\documentstyle[preprint,aps]{revtex}
\makeatletter

\newcounter{@sc}
\newcounter{@scp}
\newcounter{@t}
\newlength{\@x}
\newlength{\@xa}
\newlength{\@xb}
\newlength{\@y}
\newlength{\@ya}
\newlength{\@yb}
\newsavebox{\@pt}

\def\bezier#1(#2,#3)(#4,#5)(#6,#7){%
\c@@sc#1\relax
\c@@scp\c@@sc
\advance\c@@scp\@ne
\@xb #4\unitlength
\advance\@xb -#2\unitlength
\multiply\@xb \tw@
\@xa #6\unitlength
\advance\@xa -#2\unitlength
\advance\@xa -\@xb
\divide\@xa\c@@sc
\@yb #5\unitlength
\advance\@yb -#3\unitlength
\multiply\@yb \tw@
\@ya #7\unitlength
\advance\@ya -#3\unitlength
\advance\@ya -\@yb
\divide\@ya\c@@sc
\setbox\@pt\hbox{%
\vrule
height\@halfwidth
depth\@halfwidth
width\@wholewidth
}%
\c@@t\z@
\put(#2,#3){%
\@whilenum{%
\c@@t<\c@@scp
}%
\do{%
\@x\c@@t\@xa
\advance\@x\@xb
\divide\@x\c@@sc
\multiply\@x\c@@t
\@y\c@@t\@ya
\advance\@y\@yb
\divide\@y\c@@sc
\multiply\@y\c@@t
\raise \@y
\hbox to \z@{%
\hskip \@x\unhcopy\@pt\hss
}%
\advance\c@@t\@ne
}%
}%
}%

\makeatother
\newcommand{\sgn}{\mbox{sgn}}
\newcommand{\nonun}{\nonumber}

\newcommand{\bnor}{\left(\frac{\pi}{L}\right)^2}
\newcommand{\bnr}{\frac{\pi}{L}}

\begin{document}


\draft

\title{%
Critical Properties of \\
the Calogero-Sutherland Model with Boundaries}
\author{%
Takashi Yamamoto
}
\address{%
Yukawa Institute for Theoretical Physics,
Kyoto University, \\
Kyoto 606, Japan}
\author{%
Norio Kawakami
}
\address{%
Department of Material and Life Science,
and
Department of Applied Physics,
Osaka University, \\
Suita, Osaka 565, Japan}
\author{%
Sung-Kil Yang
}
\address{%
Institute of Physics,
University of Tsukuba, \\
Ibaraki 305, Japan}

\maketitle

\begin{abstract}
Critical properties of the Calogero-Sutherland model
of $BC_N$-type ($BC_N$-CS model) are studied.
Using the asymptotic Bethe-ansatz spectrum
of the $BC_N$-CS model,
we calculate finite-size corrections in the energy spectrum.
Since the $BC_N$-CS model does not possess
translational invariance, the finite-size spectrum
acquires the contributions coming from ``boundaries''.
We show that the low-energy critical behavior of the model
is described by $c=1$ boundary conformal field theory.
Thus the universality class of the model
is identified as a chiral Tomonaga-Luttinger liquid.
\end{abstract}

\section{Introduction}

The Calogero-Sutherland (CS) models \cite{Calogero,CS,Moser}
describe one-dimensional quantum many-body systems with
inverse-square long-rang interactions.
Among many variants of the CS model \cite{rev-CS},
a class of models which are not translationally invariant
has been known over the passed years \cite{O-Pa}.
In particular the so-called CS model of $BC_N$-type
(abbreviated as the $BC_N$-CS model hereafter)
is the most general model with $N$ interacting particles.
The $BC_N$-CS model
is intimately related to the root system of type $BC_N$
and invariant under the action of the Weyl group of type $B_N$.
Namely, the model is invariant under coordinate transformations
\begin{equation}
\label{Weyl-group}
(q_1,q_2,\cdots,q_N)
\mapsto
(\epsilon_1 q_{\sigma(1)},
\epsilon_2 q_{\sigma(2)},\cdots,
\epsilon_N q_{\sigma(N)}),
\end{equation}
where $(q_1,q_2,\cdots,q_N)\in \mbox{{\bf R}}^N$
denote the coordinates of $N$ particles,
$\epsilon_j\in\{\pm 1\}$
and $\sigma$ is an element of
the symmetric group of $N$ letters.
Roughly speaking,
the Weyl group of type $B_N$
consists of the ordinary exchange of particle coordinates
and the sign change of coordinates.
As we will see below
the latter is understood as
the mirror image of particles with respect to a boundary.

Recent works have made it clear that
the $BC_N$-CS model is relevant to
one-dimensional physics with boundaries. For instance,
it was pointed out that
the non-relativistic dynamics of quantum sine-Gordon
solitons in the presence of a boundary is described by
the $BC_N$-CS model (with $\sinh$-interaction)\cite{K-S94}.
This model is interesting in view of the quantum electric transport
in mesoscopic systems\cite{B-R,Caselle}.
The Haldane-Shastry model,
which is the discrete version of the CS model,
with open boundary conditions
can also be constructed by utilizing
the root system of type $BC_N$ \cite{S-A94,B-P-S}.
We shall present further evidence for
the relevance of the $BC_N$-CS model
to our understanding
in one-dimensional physics including boundary effects.

In this article we will analyze
the long-distance critical properties of the $BC_N$-CS model.
Since the exact energy spectrum of the model
is available \cite{Nao94},
we may apply the method
of finite-size scaling developed
in conformal field theory (CFT)
to study the critical behavior.
The same technique has already been employed
when the critical properties of the CS model
of $A_{N-1}$-type were considered \cite{K-Y91}.
The universality class of the $A_{N-1}$-CS model
is identified as a Tomonaga-Luttinger liquid
which is equivalent to $c=1$ Gaussian CFT.
In what follows we will show that,
in contrast to the $A_{N-1}$-CS model,
the $BC_N$-CS model exhibits the critical behavior described by
$c=1$ CFT with boundaries \cite{Cardy}.
Hence the universality class will be found to be
a chiral Tomonaga-Luttinger liquid \cite{ch-TLL}.

In the next section
we first introduce the $BC_N$-CS model
and review the energy spectrum of
the model obtained by using the asymptotic Bethe-ansatz.
In section 3
we consider the thermodynamic properties.
In section 4
the finite-size scaling analysis of the energy spectrum
is performed. Finally, in section 5,
we discuss various critical exponents of correlation functions.

\section{The $BC_N$-CS model}

Let us write down the Hamiltonian of the $BC_N$-CS model \cite{O-Pa}.
We put the system in finite geometry with
linear size $L$ and impose periodic boundary conditions.
The Hamiltonian is then given by
\begin{eqnarray}
{\cal H}
=
-\sum_{j=1}^N
\frac{\partial^2}{\partial q_j^2}
&+&
2\lambda(\lambda-1)
\bnor
\sum_{1\leq j<k\leq N}
\left\{
\frac{1}{\displaystyle{
\sin^2\bnr(q_j-q_k)}}
+\frac{1}{\displaystyle{
\sin^2\bnr(q_j+q_k)}}
\right\} \nonun \\
+
\lambda_1(\lambda_1 &+& 2\lambda_2-1)
\bnor
\sum_{j=1}^N \frac{1}{\displaystyle{
\sin^2\bnr q_j}}
+
4\lambda_2(\lambda_2-1)
\bnor
\sum_{j=1}^N \frac{1}{\displaystyle{
\sin^2\bnr 2q_j}},
  \label{b-tri-hamiltonian1}
\end{eqnarray}
where
$\lambda, \,  \lambda_1$ and $\lambda_2$
are coupling constants
which are assumed to be non-negative.
It is clearly seen that the Hamiltonian
(\ref{b-tri-hamiltonian1}) is invariant
under the action (\ref{Weyl-group})
of the Weyl group of type $B_N$.
There exist several interaction terms
which will need explanation.
The term
$1/\sin^2(\pi/L)(q_j+q_k)$
expresses the two-body interaction
between the $j$-th particle
and
the ``mirror-image'' (we place a mirror at
the origin $q=0$)
of the $k$-th particle ($j\ne k$).
The term
$1/\sin^2(\pi/L)q_j^2$
may be interpreted as
the potential due to {\it impurity}
located at  the origin.
The term
$1/\sin^2(\pi/L)2q_j$ describes
the interaction between
the $j$-th particle
and its own ``mirror-image''.
All these terms
required by
invariance
under the
action of the Weyl group
of type $B_N$
violate translational invariance.
Therefore,
the total momentum
is not a good quantum number for the $BC_N$-CS model.

The Hamiltonian (\ref{b-tri-hamiltonian1})
can be cast into another form just by using
the elementary identity
$\sin 2A=2\sin A\cos A$.
One gets
\begin{eqnarray}
{\cal H}
=
-\sum_{j=1}^N
\frac{\partial^2}{\partial q_j^2}
&+&
2\lambda(\lambda-1)
\bnor
\sum_{1\leq j<k\leq N}
\left\{
\frac{1}{\displaystyle{
\sin^2\bnr(q_j-q_k)}}
+\frac{1}{\displaystyle{
\sin^2\bnr(q_j+q_k)}}
\right\}
\nonun
\\
&+&
\mu(\mu-1)
\bnor
\sum_{j=1}^N \frac{1}{\displaystyle{
\sin^2\bnr q_j}}
+
\nu(\nu-1)
\bnor
\sum_{j=1}^N \frac{1}{\displaystyle{
\cos^2\bnr q_j}},
 \label{b-tri-hamiltonian2}
\end{eqnarray}
where
$\mu=\lambda_1+\lambda_2$, $\nu=\lambda_2$.
In this form of the Hamiltonian
the term $1/\sin^2(\pi/L)(q_j+q_k)$ is regarded
as the boundary potential as before, while the last two terms in
(\ref{b-tri-hamiltonian2}) are regarded as the impurity potentials with
the strength determined by $\mu$ and $\nu$ respectively.
The Hamiltonian (\ref{b-tri-hamiltonian2}) is suitable for our
present considerations.

The eigenvalues of the Hamiltonian
(\ref{b-tri-hamiltonian2}) of the
$BC_N$-CS model
have been obtained by
one of the authors\cite{Nao94}\footnote{%
Precisely speaking,
this reference treated the case
with $\nu=0$ (the $B_N$-CS model).
However, we can easily obtain the formula for
the $BC_N$-CS model. The spectrum was also derived in
\cite{B-P-S}.}.
The energy spectrum so obtained is
shown to be reproduced exactly with
the use of the
asymptotic Bethe-ansatz (ABA) method
\cite{Nao94}.
Let us recall the ABA formula for the $BC_N$-CS
model. First of all the total energy of the system
takes the form
\begin{equation}
\label{one-exc-ene}
E_N
=
\sum_{j=1}^N
{k_j}^2,
\end{equation}
where pseudomomenta $k_j$'s
satisfy
$k_1>k_2>\cdots>k_N>0$ and obey
the ABA equations
\begin{eqnarray}
\label{exci-rapidi}
k_j L
&=&
2\pi I_j
+
\pi
(\lambda-1)
\sum_{l=1,l\ne j}^N
\left\{
\sgn(k_j-k_l)
+
\sgn(k_j+k_l)
\right\}
\nonumber
\\
& &
+
\pi
(\mu-1)
\sgn(k_j)
+
\pi
(\nu-1)
\sgn(k_j),\hskip10mm j=1,\cdots,N,
\end{eqnarray}
with $\sgn(x)=1$ for $x>0$, $=0$ for $x=0$ and $=-1$ for $x<0$.
Here
$I_j\ (j=1,\cdots,N)$ are
positive integers
with $I_1>I_2>\cdots>I_N>0$.
These are quantum numbers
which
characterize the excited states.

We emphasize here that, in contrast to the
$A_{N-1}$-CS model, the Fermi surface of the
$BC_N$-CS model consists of a single point.
This is due to the fact that
pseudomomenta $k_j$ which are
solutions to (\ref{exci-rapidi})
are distributed only over the semi-infinite region
as is shown in Fig.1a.
Therefore,
in view of the bosonization picture,
it implies that the
low-energy critical behavior
of the $BC_N$-CS model
will be effectively
described by
a left (or right)-moving sector of
CFT (see Fig.1b).
In addition to this, we also notice that
the form of our Bethe-ansatz
equations (\ref{exci-rapidi}) is quite close to that
appeared in the studies
of the nonlinear Schr\"{o}dinger equation
on the half line\cite{gaudin71,B-M88}
as well as the $XXZ$ model with
open boundary conditions \cite{H-Q-B87,A-B-B-B-Q87}.
The critical behavior observed in these
models \cite{gaudin71,B-M88,H-Q-B87,A-B-B-B-Q87}
is well described by boundary CFT \cite{Cardy}.
It is inferred from these points that boundary CFT
will play a role in our study of the $BC_N$-CS model.

{}Finally we rewrite
our ABA equation (\ref{exci-rapidi}) for further convenience.
As has already been mentioned,
all the pseudomomenta $k_j$ are positive.
However, one can make a trick so that
$k_j$ takes values in $(-\infty, \infty)$
as in the
bulk system.
To realize this let us define
$I_{-j}=-I_j,\,  I_0=0,\, k_{-j}=-k_j$ and $k_0=0$
with $j=1,\cdots,N$,
then we have
\begin{eqnarray}
\label{extent}
k_j
&=&
4\pi\frac{1}{2L}I_j
+
2\pi(\lambda-1)
\frac{1}{2L}
\sum_{l=-N}^N
\sgn(k_j-k_l)
\nonumber
\\
& &
+
\frac{\pi}{L}(\mu+\nu-2)\sgn(k_j)
-
\frac{\pi}{L}(\lambda-1)\sgn(2k_j)
-
\frac{\pi}{L}(\lambda-1)\sgn(k_j),
\end{eqnarray}
where $j=-N,-N+1,\cdots,N$.
The last two terms
in (\ref{extent})
arise since
the summation in (\ref{exci-rapidi})
does not include the terms
$l=j$ and $l=0$.
Now the system
turns out to have
linear size
$2L$ and the number of particles
becomes $2N+1$.
Note that the density of the system
does not change.
This doubling trick is
known to be efficient when studying
one-dimensional physics
with boundaries \cite{gaudin71,B-M88,H-Q-B87,A-B-B-B-Q87}.

\section{Thermodynamic Properties}

The purpose in this section is to discuss
thermodynamics of the $BC_N$-CS model.
Let us first consider the
system at zero temperature.
All the states inside of the interval
$[-k_F,k_F]$ are occupied,
where the Fermi momentum $k_F$
is defined as $k_F=\mbox{max}\{k_j\}$.
The thermodynamic limit is taken by
$2L\rightarrow \infty,\ 2N+1\rightarrow \infty$
with the density $(2N+1)/2L$ fixed.
As usual we define the
density of states
by
\begin{equation}
\label{d-s}
\lim_{L\mapsto \infty}
\frac{1}{2L(k_j-k_{j+1})}
=\rho(k),
\end{equation}
and the sum is converted into integral
\begin{equation}
\label{integral}
\frac{1}{2L}\sum_{j=-N}^N (\ )
\mapsto
\int_{-k_F}^{k_F}dk\rho(k) (\ ).
\end{equation}

{}From
(\ref{extent}), (\ref{d-s}), (\ref{integral})
and
$\frac{d}{d x}\sgn(x)
=2\delta(x)$,
it is shown that
\begin{eqnarray}
  \label{twice-aba-eq}
1
=
4\pi\rho(k)
&+&
4\pi(\lambda-1)
\int_{-k_F}^{k_F}
dk'\delta(k-k')\rho(k')
+
\frac{2\pi}{L}(\mu+\nu-2\lambda)\delta(k),
\end{eqnarray}
where the boundary effect
manifests itself
in the last term ($\sim 1/L$).
Notice that even for
$\mu=\nu=0$,
it still modifies the equation.
Upon taking the thermodynamic limit one can neglect the boundary
term.
The resulting
equation is the same as for the
$A_{N-1}$-CS model \cite{CS}.
Then it is immediate to get
\begin{eqnarray}
  \label{density}
\rho(k)
&=&
\frac{1}{4\pi \lambda},
\\
\label{fermi-mo}
k_F
&=&
2\pi\lambda d,
\end{eqnarray}
where we have put $d=N/L$.
It is also straightforward to compute
the ground-state energy,
\begin{eqnarray}
  \label{energy-aba}
E^{(0)}
=
\sum_{j=-N}^{N}(k_j^{(0)})^2
=
2L
\int_{-k_F}^{k_F}
dkk^2\rho(k)
=
2L\cdot \epsilon^{(0)}
\end{eqnarray}
with $\epsilon^{(0)}=4
\pi^2\lambda^2d^3/3$
in the $2L \rightarrow \infty$ limit.

It is not difficult to extend the above analysis to the
finite temperature case.
At finite temperatures the
pseudomomenta distribute over the infinite region
$(-\infty,\infty)$.
One finds
\begin{eqnarray}
  \label{twice-aba-eq-finite}
1
=
4\pi(\rho(k)+\rho^h(k))
&+&
4\pi(\lambda-1)
\int_{-\infty}^{\infty}
dk'\delta(k-k')\rho(k')
\nonumber
\\
&+&
\frac{2\pi}{L}(\mu+\nu-2\lambda)\delta(k),
\end{eqnarray}
where
$\rho^h(k)$
is the hole density.
Let $2L\rightarrow \infty$, then we have
\begin{equation}
\label{finite-tba}
\rho(k)+\frac{1}{\lambda}\rho^h(k)
=
\frac{1}{4\pi \lambda}.
\end{equation}Following now the familiar procedure,
we obtain the thermodynamic Bethe-ansatz equation,
\begin{equation}
\label{tba}
\epsilon(k)
=
k^2-\mu_c
+
(\lambda-1)T
\log
\left\{
1
+
\exp
\left(
-
\frac{1}{T}
\epsilon(k)
\right)
\right\},
\end{equation}
where $T$ is the temperature,
$\mu_c$
is the chemical potential and the energy
density $\epsilon(k)$ of particles
is defined by
\begin{equation}
\label{p-p-d}
\frac{\rho(k)}{\rho^h(k)}
=
\exp
\left(
-\frac{1}{T}
\epsilon(k)
\right).
\end{equation}
Performing the low-temperature expansion
of the free energy $F(T)$ which is given by
\begin{equation}
(F(T)-\mu_c(2N+1))/(2L)
=
-
\frac{T}{4\pi}\int_{-\infty}^\infty
dk \log(1+e^{-\frac{1}{T}\epsilon(k)}),
\end{equation}
we have
\begin{equation}
\label{lte}
 F(T)
\simeq
 F(T=0)
-
\frac{\pi T^2}{6(4\pi \lambda d)}.
\end{equation}

The second term in (\ref{lte}) is responsible for the linear specific heat
$C$ as $T\rightarrow 0$.
It is well recognized that the coefficient in $C$
is universal modulo the Fermi velocity $v_{\mbox{{\tiny F}}}$
which is not universal \cite{B-C-N}.
In translationally invariant systems the Fermi velocity is
determined by the dispersion relation.
In the $BC_N$-CS model, however, one cannot rely on
the dispersion relation since the momentum
is not a good
quantum number.
So, in order to determine $v_{\mbox{{\tiny F}}}$,
we have to take another point of view.
As we observed, eqs.(\ref{density}), (\ref{fermi-mo}) and (\ref{finite-tba})
coincide with those obtained in the $A_{N-1}$-CS model.
Hence we may regard the $A_{N-1}$-CS model as the bulk
counterpart of the $BC_N$-CS model.
Since the $A_{N-1}$-CS model is described in terms of $c=1$ CFT \cite{K-Y91}
we assume that the central charge for the $BC_N$-CS model
is also given by $c=1$. Then, comparing $C$ obtained from (\ref{lte})
to the formula $C=\pi cT/(3v_{\mbox{{\tiny F}}})$ \cite{B-C-N}
with $c=1$ we find $v_{\mbox{{\tiny F}}}=4\pi\lambda d$.
We shall see in section 5 that the finite-size spectrum is in fact in accord
with $c=1$ CFT.

\section{Finite-size scaling analysis}

In this section we perform the finite-size
scaling analysis of the energy spectrum of the $BC_N$-CS model.
To begin with, we summarize several fundamental formulas in boundary CFT
\cite{Cardy} which we will need to analyze the energy spectrum.
Let us first recapitulate the finite-size
scaling form of the ground-state energy predicted by
conformal invariance under {\it free boundary conditions}
\cite{B-C-N}
\begin{equation}
E^{(0)}
=
L\epsilon^{(0)} + 2f - \frac{\pi v_{\mbox{{\tiny F}}}}{24L}c \ ,
\label{fsscft}
\end{equation}
where $\epsilon^{(0)}$ and $f$ are, respectively,
the bulk limits of the ground-state energy density
and the boundary energy,
$v_{\mbox{{\tiny F}}}$ is the velocity of the elementary excitations.
The Virasoro central charge $c$
which specifies the universality class of the system
appears as the universal amplitude of the $1/L$ term in (\ref{fsscft}).

{}From the scaling behavior of the excitation energy one can read off
the boundary critical exponents $x_b$ \cite{Cardy}.
This exponent $x_b$ governs the power-law decay
(parallel to the boundary surface) of a two-point function.
Suppose a critical system on the half-plane
$\{(y,\tau)\in
\mbox{{\bf R}}_{\geq 0}\times \mbox{{\bf R}}\}$
with a surface at $y=0$.
($y$ is the perpendicular distance from a point $(y,\tau)$
to the boundary
and $\tau$ means the imaginary time.)
Let ${\cal O}(y,\tau)$ be a local operator.
We consider its two-point correlation function
$G(y_1,y_2,\tau)=
\langle{\cal O}(y_1,\tau_1){\cal O}(y_2,\tau_2)\rangle$,
which is a function of $\tau=\tau_1-\tau_2$ because
of translational invariance along the surface. For
$|\tau|\gg y_1,\, y_2$, we obtain
the asymptotic form of $G$,
\begin{equation}
\label{long-time-asymp}
G(y_1,y_2,\tau)
\sim
\frac{1}{\tau^{2x_b}}.
\end{equation}

To evaluate $x_b$ we have to examine the scaling law
\begin{equation}
E-E^{(0)}
= \frac{\pi v_{\mbox{{\tiny F}}}}{L}x_b
\label{ex-corr}
\end{equation}
with $E$ being the excitation energy.
It usually happens that the value of $x_b$ is distinct from that of
the bulk exponent for certain scaling operator.
In terms of CFT, the bulk exponent is expressed as
the sum of left and right conformal weights,
while the boundary exponent is equal to the left (or right)
conformal weight.

Let us now turn to the $BC_N$-CS model.
It is convenient to manipulate the ABA equations
(\ref{exci-rapidi}) directly.
We can easily solve (\ref{exci-rapidi})
to obtain
\begin{eqnarray}
k_j
&=&
\frac{2\pi}{L}
\left[
I_j
-
\left(
N-j+1
\right)
\right]
+
k_j^{(0)}, \hskip10mm j=1,\cdots,N,
\end{eqnarray}
where
\begin{equation}
\label{gra-rapidi}
k_j^{(0)}
=
\frac{2\pi}{L}
\left[
\lambda(N-j)+\frac{\mu+\nu}{2}
\right].
\end{equation}
The ground state is thus specified by the quantum numbers
$I_j^{(0)}=N-j+1,\ (j=1,\cdots,N)$, from which
we get the Fermi point $I_1^{(0)}=N$
and the Fermi momentum
$k_F=2\pi\lambda N/L+\pi(\mu+\nu-2\lambda)/L$.
The ground-state energy is then obtained as
\begin{eqnarray}
\label{one-gro-ene}
E_N^{(0)}
&=&
\sum_{j=1}^N
\left(
k_j^{(0)}
\right)^2
\nonumber
\\
&=&
\left(
\frac{2\pi}{L}
\right)^2
\left[
\frac{1}{3}
\lambda N
+
\frac{1}{2}
\lambda(\mu+\nu-\lambda)N^2
+
\frac{1}{12}
\left(
3(\mu+\nu-\lambda)^2-\lambda^2
\right)N^3
\right].
\end{eqnarray}
We make a power expansion of (\ref{one-gro-ene}) with respect to
$1/L$ while keeping the particle density
$d=N/L$ fixed.
The result reads
\begin{equation}
\label{fss-gro}
E_N^{(0)}
=
\epsilon^{(0)}L
+
2f
+
\frac{\pi v_{\mbox{{\tiny F}}}}{L}
\lambda(\Delta N_b)^2
-
\frac{\pi v_{\mbox{{\tiny F}}}}{12L}\lambda,
\end{equation}
where $f=
\pi^2
\lambda
(\mu+\nu-\lambda)d^2$ and
\begin{eqnarray}
\Delta N_b
=
\frac{\mu+\nu-\lambda}{2\lambda}.
\end{eqnarray}
In (\ref{fss-gro})
there appear no higher-order terms with $L^{-m}(m\geq 2)$.
Note also the symmetric dependence
of $f$ and $\Delta N_b$ on $\mu, \, \nu$.

There are several points
which should be noticed in (\ref{fss-gro}). First of all,
besides the thermodynamic energy density $\epsilon^{(0)}$
already computed in (\ref{energy-aba}), one finds
the {\it boundary energy} $2f$
in the term of order $L^0$, which is due to the absence of
translational invariance in the system.
The next order corrections proportional to $1/L$ turn out to provide
valuable information on "boundary effects".
To see this, let us proceed a bit carefully by having decomposed the
$1/L$-contributions into the last two terms in (\ref{fss-gro}).
We first recall that the size-dependence of the interaction
is inevitably  introduced for $1/r^2$ systems, as
seen in (\ref{b-tri-hamiltonian2}),
when dealing with interacting particles in finite geometry.
This gives rise to nonuniversal
$1/L$-corrections to the ground-state energy in addition to the universal one,
as observed in the $A_{N-1}$-CS model \cite{K-Y91}.
In (\ref{fss-gro}), therefore, we think that
the term $-\pi v_{\mbox{{\tiny F}}} \lambda /(12L)$ suffers from such
nonuniversal contaminations which,
in direct comparison with (\ref{fsscft}), yield the wrong
value for the central charge.

The other $1/L$-correction term, $\pi v_{\mbox{{\tiny F}}}
\lambda (\Delta N_b)^2/L$, is
more interesting and understood as the
"boundary effect" which consists of
two kinds of contributions.
As seen from (\ref{extent}), when we convert the $BC_N$ system to
the {chiral} system by using a trick of mirror image,
we are left with particles moving only in one
direction feeling the {\it boundary potential}
depending on $\lambda$, in addition to
the {\it impurity potential} depending on $\mu$ and $\nu$.
These two types of scattering effects
are combined into a quadratic form with respect to
the "fractional quantum number" $\Delta N_b$
depending on both $\mu+\nu$ and $\lambda$.  Note that
the quantum number $\Delta N_b$ physically represents the
phase shift due to the scattering by the impurity- and
boundary-potentials.
Thus our ground-state energy $E_N^{(0)}$ is considered as
the phase-shifted ground-state energy\cite{A-L94}.
If we imagine a hypothetical system which does not include these boundary
contributions, the corresponding ground-state
energy $\tilde{E}_N^{(0)}$ is written as
\begin{equation}
\label{new-b-e}
\tilde{E}_N^{(0)} = E_N^{(0)} - \frac{2\pi v_{\mbox{{\tiny F}}}}{L}
\frac{\lambda}{2}(\Delta N_b)^2.
\end{equation}

Having discussed the ground-state energy in detail,
we next wish to calculate the finite-size corrections to the
excited states. Looking at the ABA equations (\ref{exci-rapidi})
let us create an excited state by adding $\Delta N$ particles to
the ground-state configuration.  In this case,
we have the pseudomomenta
\begin{equation}
\label{p-n-c}
k_j= \frac{2\pi}{L}
\left[ \lambda(N+\Delta N-j) +\frac{\mu+\nu}{2}
\right],
\end{equation}from which we immediately obtain
the finite-size corrections to leading order in
$1/L$,
\begin{eqnarray}
\label{nun-dis}
E_{N+\Delta N}^{(0)}-E_{N}^{(0)} &\simeq&
\mu_c^{(0)} \Delta N + \frac{\pi}{L}
\left[ 4\pi\lambda(\mu+\nu-\lambda)d\Delta N
+ 4\pi\lambda^2d(\Delta N)^2\right]
\nonumber
\\
&=&
\mu_c^{(0)} \Delta N  + \frac{\pi v_{\mbox{{\tiny F}}}}{L} \lambda
\left( \Delta N + \Delta N_b\right)^2
- \frac{\pi v_{\mbox{{\tiny F}}}}{L}
\lambda(\Delta N_b)^2,
\end{eqnarray}
where $\mu_c^{(0)} =\partial \epsilon^{(0)}/\partial d
={k_F}^2$
is the chemical potential. Note that this expression for the finite-size
spectrum is essentially the same as that derived for the charge sector in the
Kondo problem (see (49) in \cite{fky}). If we redefine $E_N^{(0)}$ by
$E_N^{(0)}-\mu_c^{(0)} N$, we find
\begin{equation}
\label{s-fss}
E_{N+\Delta N}^{(0)} - \tilde{E}_N^{(0)}
= \frac{2\pi v_{\mbox{{\tiny F}}}}{L}
\frac{\lambda}{2} \left(\Delta N + \Delta N_b \right)^2.
\end{equation}
Since any excitations which carry currents with
large momentum transfer are barred
due to the absence of translational invariance
in the $BC_N$-CS model, the remaining possible
type of low-energy excitations are provided by
particle-hole excitations labeled by non-negative integers
$n$.
The corresponding energy is simply obtained by adding
$2\pi v_{\mbox{{\tiny F}}}n/L$ to (\ref{s-fss}).
Hence we have
\begin{equation}
\label{s-fss-all}
E - \tilde{E}_N^{(0)}
=
\frac{2\pi v_{\mbox{{\tiny F}}}}{L}
\left[
\frac{\lambda}{2}
\left(\Delta N + \Delta N_b \right)^2+n \right],
\end{equation}
where $E$ denotes the energy of the excited state
specified by $(\Delta N,\Delta N_b,n)$.
In the next section we argue that our result (\ref{s-fss-all})
is in accordance with the scaling law in $c=1$ boundary CFT.

\section{Correlation functions}

Now that we have evaluated the finite-size corrections it is possible to
read off various critical exponents using the scaling relation (\ref{ex-corr}).
When comparing our result (\ref{s-fss-all})
with (\ref{ex-corr}) we have to replace $L$ with
$2L$ since $L$ has been defined as the periodic length of the system.
Bearing this in mind let us take an operator $\psi_b$
which corresponds to the phase-shifted ground state.
This operator can be assumed to be the boundary changing
operator\cite{A-L94}. With this point of view, the phase-shifted ground state
is an excited state relative to $\tilde E^{(0)}_N$ in (\ref{new-b-e}).
The scaling
dimension of $\psi_b$ is obtained as
\begin{equation}
x_{\psi_b}
=
\frac{L}{\pi v_{\mbox{{\tiny F}}}}
\left(
E_N^{(0)}
-
\tilde{E}_N^{(0)}
\right)
=
\frac{1}
{2\xi^2}
\left(
\Delta N_b
\right)^2,
  \label{bound-ex-gra}
\end{equation}
where we have put
$\xi=1/\sqrt{\lambda},\ \zeta=1/\sqrt{\mu+\nu}$, and hence
$\Delta N_b=(\xi^2-\zeta^2)/(2\zeta^2)$.

We next consider an operator
$\phi$ which induces the particle number change as well as the
particle-hole excitation
in the phase-shifted ground state. From (\ref{s-fss})
and (\ref{ex-corr}) we have
\begin{equation}
x_{\phi}
=
\frac{L}{\pi v_{\mbox{{\tiny F}}}}
\left(
E_{N+\Delta N}^{(0)}
-
\tilde{E}_N^{(0)}
\right)
=
\frac{1}
{2\xi^2}
\left(
\widehat{\Delta N}
\right)^2+n,
  \label{bound-ex}
\end{equation}
where
\begin{equation}
\label{modi-q-n}
\widehat{\Delta N}
=
\Delta N
+
\Delta N_b.
\end{equation}
Scaling dimensions
(\ref{bound-ex-gra}) and
(\ref{bound-ex}) take the form of conformal weights characteristic of
$c=1$ CFT. The radius $R$ of compactified $c=1$ free boson is taken to be
$R=\xi$. Let us concentrate on the self-dual point $R=1/\sqrt{2}$
({\it i.e.} $\lambda =2$) where the symmetry is known to be enhanced to
the level-1 $SU(2)$ Kac-Moody algebra. In the $BC_N$-CS model we have
the other continuous parameters $\mu, \, \nu$ which should also be tuned
to achieve the $SU(2)$ point. It turns out that $\mu +\nu =0,\, 1,\, 2,\, 3$
and $4$ with $\lambda =2$ are the desired points.
This follows from the following observations:
When $\mu +\nu =2$ we have $\Delta N_b=0$ and hence
\begin{equation}
x_{\phi}=\frac{1}{4} (2\Delta N)^2+n
\end{equation}
which is the conformal weight for the spin-0 irreducible representation
of the level-1 $SU(2)$ Kac-Moody algebra. When $\mu +\nu =4$ or $0$ we get
$\Delta N_b=\pm 1/2$ and thus
\begin{equation}
x_{\phi}=\frac{1}{4} (2\Delta N +1)^2+n
\end{equation}
which is the conformal weight of spin-$1/2$ irreducible representation.
When $\mu +\nu =3$ or $1$ we have $\Delta N_b=\pm 1/4$, thereby
\begin{equation}
x_{\phi}=\frac{1}{16} (4\Delta N +1)^2+n.
\end{equation}
This is the conformal weight for the unique irreducible representation
of the level-1 twisted $SU(2)$ Kac-Moody algebra \cite{twist}.
The highest-weight state with $x_\phi =1/16$ is a twist field in $c=1$
CFT. Several
$SU(2)$ points identified in \cite{B-P-S} are in agreement with our result.
Thus we conclude that
the low-energy critical behavior of the $BC_N$-CS model
is described in terms of $c=1$ boundary CFT,
{\it i.e.} the universality class of a chiral Tomonaga-Luttinger liquid.

{}Further considerations on the low-energy critical properties
of the $BC_N$-CS model require a clear distinction between two pictures
corresponding to two possible sets of quantum numbers. One is a set
of quantum numbers $(\Delta N,\Delta N_b,n)$ and the other is a set of
$(\widehat{\Delta N},\  n)$ where $\widehat{\Delta N}$ is regarded as
the ordinary particle number change in (\ref{bound-ex})
(forgetting about $\Delta N_b$ in
(\ref{modi-q-n})). The picture based on the set
$(\Delta N,\Delta N_b,n)$ is relevant when describing
the long-time asymptotic behavior
of the system in which we suddenly turn on the boundary effects
in the ground state. The X-ray
absorption singularity in the Kondo problem, for instance,
is considered in this type of picture \cite{A-L94,fky}.
The boundary changing operator $\psi_b$ is described
in this picture with $(\Delta N,\Delta N_b,n)=(0,\Delta N_b,0)$.
If we use the set $(\widehat{\Delta N},\  n)$ instead,
our picture is independent of $\zeta$
and adequate to compute the critical exponents of
ordinary correlation functions with boundary effects.

Let us consider the one-particle Green function
in the above two pictures. Let $(\Delta N,\Delta N_b,n)
=(1,\Delta N_b,0)$ in the first picture.
This choice of quantum numbers determines the
long-time asymptotic behavior of
the field correlator (the one-particle Green function)
when boundary potentials are turned on at $\tau=0$,
\begin{equation}
\label{scaling}
\langle \Psi^{\dag}(\tau)\Psi(0) \rangle_{{\rm sudden}}
\sim
\frac{1}{\tau^{2x_{G}}},
\end{equation}
where
\begin{equation}
\label{s-d}x_{G}
= \frac{1}{2\xi^2}(1+\Delta N_b)^2
= \frac{1}{8\xi^2}
\left(
1
+
\frac{\xi^2}{\zeta^2}
\right)^2.
\end{equation}
Here $\langle \cdots \rangle_{{\rm sudden}}$ stands for the expectation value
when the boundary potential is suddenly
switched on.
On the other hand, if we let $(\widehat{\Delta N},n)
=(1,0)$ in the second picture,
the field correlator takes the form,
\begin{equation}
\label{scaling2}
\langle \Psi^{\dag}(\tau)\Psi(0) \rangle
\sim
\frac{1}{\tau^{2x_{g}}},
\end{equation}
where
\begin{equation}
\label{s-d2}
x_{g}
=
\frac{1}{2\xi^2},
\end{equation}
which describes the ordinary one-particle Green function.
In this case, the boundary critical exponent $x_{g}$
linearly depends on $\lambda$.
Contrary to these Green functions,
the density-density correlation function
is controlled by the excitations which do not change
the number of particles.  Hence, it should have
the long-time asymptotic form,
\begin{equation}
\langle \rho(\tau)\rho(0) \rangle
\sim
\frac{1}{\tau^{2}},
\end{equation}
which follows by taking the quantum number $(\widehat{\Delta N},n)=(0,1)$
in (33).
Note that there do not appear anomalous exponents
in this correlator.  One can easily see that this is also the case for
sub-leading terms $\tau^{-2k}$ in which
the quantum number is chosen as $(\widehat{\Delta N},n)=(0,k)$.
This fact will be confirmed shortly in the following.

We  now compare our result with the explicit calculations of
the dynamical correlation function.
In the case $\lambda=1,\, \nu=0$ with $\mu$ arbitrary
which corresponds to the noninteracting system,
the dynamical density-density correlation function for
the $BC_N$-CS model
has been obtained by Mac\^{e}do \cite{Macedo}
(see also \cite{MIT}).
In the thermodynamic limit, the density-density correlation function
$G(y_1,y_2,\tau)$ has the form
\begin{eqnarray}
\label{auto-corr}
G(y_1,y_2,\tau)
=
\frac{\pi^4}{4}
y_1 y_2
&&
\int_1^\infty d u_1
e^{-\frac{1}{2}\pi^2\tau u_1}
J_{\mu-\frac{1}{2}}(\pi y_1 \sqrt{u_1})
J_{\mu-\frac{1}{2}}(\pi y_2 \sqrt{u_1})
\nonumber
\\
\times
&&
\int_0^1 d u_2
e^{\frac{1}{2}\pi^2 \tau u_2}
J_{\mu-\frac{1}{2}}(\pi y_1 \sqrt{u_2})
J_{\mu-\frac{1}{2}}(\pi y_2 \sqrt{u_2}),
\end{eqnarray}
where $J_{\nu}(z)$ is the Bessel function and
$\tau$ is the imaginary time.
When $\mu=1/2+m\ (m=0,1,\cdots)$ it is not difficult to evaluate the
large-$\tau$ asymptotic behavior by making use of the series expansion of
$J_m(z)$. After some algebra we obtain
\begin{equation}
\label{special-case}
G(y_1,y_2,\tau)
=
\sum_{k=1}^\infty
A_k(y_j)
\left(
\frac{1}{\tau}
\right)^{2k}
+
\sum_{l=0}^\infty
B_l(y_j)
\left(
\frac{1}{\tau}
\right)^{l+m+2}
e^{-\frac{1}{2}\pi^2\tau},
\end{equation}
where
$A_k(y_j), B_l(y_j)$ are some functions.
As $\tau\rightarrow \infty$ with $y_1,\, y_2$ fixed, the second term vanishes
exponentially, yielding
\begin{equation}
\label{asymp-kappa}
G(y_1,y_2,\tau) \simeq \frac{A_1}{\tau^2}+\frac{A_2}{\tau^4}
+\frac{A_3}{\tau^6}+\cdots .
\end{equation}
Notice that the exponents are independent of
$m$  ({\it i.e.}, $\mu$).
The density-density correlation function is
considered in the picture based on $(\widehat{\Delta N},\ n)$.
Then we see from (\ref{bound-ex}) that all these exponents are
precisely understood in terms of the excitations
$(\widehat{\Delta N},\ n)=(0,k)$.
This means that the correlation function $G$
is dominated by the particle-hole excitations,
and hence there
is no way of depending on $\lambda$.
Therefore the result (\ref{asymp-kappa})
completely agrees with our prediction
by CFT analysis. We are thus led to conclude that
the power-law decay in (\ref{asymp-kappa}) is universal irrespective of
$\lambda$ (but with $\nu =0$ fixed)
though (\ref{asymp-kappa}) is verified at $\lambda =1$.
We stress that this remarkable feature in the density-density
correlation function is inherent in {\it chiral}
Tomonaga-Luttinger liquids \cite{ch-TLL}.

{}Finally we briefly mention possible applications to
the (chiral) random matrix theory\cite{Mehta91}.
Let us recall the $B_N$ Calogero-Moser model ($B_N$-CM model)
in the rational form \cite{O-Pa},
\begin{eqnarray}
\label{cm-model}
{\cal H}_{\mbox{{\tiny C-M}}}
=
-\sum_{j=1}^N
\frac{\partial^2}{\partial x_j^2}
&+&
2\lambda(\lambda-1)
\sum_{1\leq j<k\leq N}
\left\{
\frac{1}{(x_j-x_k)^2}
+
\frac{1}{(x_j+x_k)^2}
\right\}
\nonumber
\\
&+&
\mu(\mu-1)
\sum_{j=1}^N
\frac{1}{x_j^2}
+
\omega^2
\sum_{j=1}^N x_j^2,
\end{eqnarray}
with $\omega>0$. In the thermodynamic limit,
this model belongs to the same universality class as the $B_N$-CS model
which is equivalent to the $BC_N$-CS model at $\nu =0$.
The ground-state wave function for the $B_N$-CM model takes the form of
Jastrow-type \cite{O-Pa}
\begin{equation}
\label{cm-gr-st}
\Psi^{(0)}
(x_1,x_2,\cdots,x_N)
=
{\cal N} \prod_{1\leq j<k\leq N}
|x_j^2-x_k^2|^{\lambda}
\prod_{l=1}^N
|x_l^2|^{\frac{\mu}{2}}
\exp
\left(
-\frac{1}{2}\omega x_l^2
\right),
\end{equation}
where ${\cal N}$ is a calculable normalization constant.
Notice that
$\Psi^{(0)}
(x_1,x_2,\cdots,x_N)$
depends only on the $x_j^2$'s.
Then, introducing new variables $z_j=x_j^2$, one should note that
$|\Psi^{(0)} (x_1,x_2,\cdots,x_N)|^2$ is identical to the probability
distribution function for the eigenvalues $z_j$ of the
Laguerre ensemble when $\lambda=1/2,1$ and $2$
(with appropriate values of
$\mu$ and $\omega$) corresponding to the ensembles of orthogonal,
unitary and symplectic types \cite{Mehta91},
respectively.
Therefore, it will be very interesting if
the long-time asymptotic behavior of
correlation functions in the $B_N$-CM model obtained in the present work
is directly compared with the
results in the Laguerre random matrix theory.


In summary, we have investigated
boundary critical phenomena in the $BC_N$-CS model.
The boundary effects come from both
the impurity potentials and interactions
between particles and ``image'' particles.
Making use of boundary CFT, we have obtained boundary critical
exponents, and clarified the
critical properties of the $BC_N$-CS model
in terms of chiral Tomonaga-Luttinger liquids.

\vskip10mm


{\bf Acknowledgements}\\
T.Y. was supported by the Yukawa memorial foundation
and the COE (Center of Excellence) researchers program
of the Ministry of Education, Science and Culture, Japan.
N. K. was partly supported by a Grant-in-Aid from the Ministry
of Education, Science and Culture, Japan.
The work of S.-K.Y. was supported in part by Grant-in-Aid
for Scientific Research on Priority Area 231 ``Infinite Analysis'',
the Ministry of Education, Science and Culture, Japan.



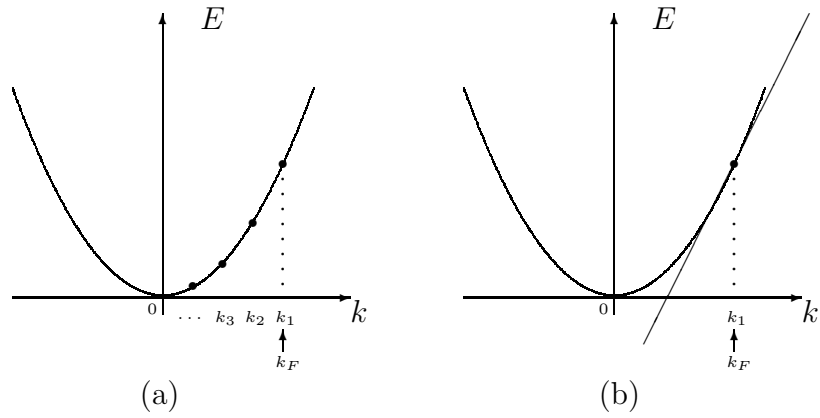
\begin{figure}
\vspace*{3cm}
\begin{picture}(16,6)(0,0)
\put(4,2.5){\vector(0,1){4}}
\put(4.5,6.3){$E$}
\put(2,2.72){\vector(1,0){4.5}}
\put(6.5,2.4){$k$}
\put(5.6,4.5){\circle*{0.1}}
\multiput(5.6,4.3)(0,-0.2){8}{\circle*{0.01}}
\put(5.5,2.4){{\tiny \mbox{$k_1$}}}
\put(5.6,2.0){\vector(0,1){0.3}}
\put(5.5,1.8){{\tiny \mbox{$k_F$}}}
\put(5.2,3.71){\circle*{0.1}}
\put(5.1,2.4){{\tiny \mbox{$k_2$}}}
\put(4.8,3.17){\circle*{0.1}}
\put(4.7,2.4){{\tiny \mbox{$k_3$}}}
\put(4.4,2.88){\circle*{0.1}}
\put(4.2,2.4){{\tiny \mbox{$\cdots$}}}
\put(3.8,2.5){{\tiny \mbox{$0$}}}
\bezier{500}(2,5.5)(4,0)(6,5.5)
\put(3.7,1.3){(a)}
\put(10,2.5){\vector(0,1){4}}
\put(10.5,6.3){$E$}
\put(8,2.72){\vector(1,0){4.5}}
\put(12.5,2.4){$k$}
\put(11.6,4.5){\line(1,2){1}}
\put(11.6,4.5){\line(-1,-2){1.2}}
\put(11.6,4.5){\circle*{0.1}}
\multiput(11.6,4.3)(0,-0.2){8}{\circle*{0.01}}
\put(11.5,2.4){{\tiny \mbox{$k_1$}}}
\put(11.6,2.0){\vector(0,1){0.3}}
\put(11.5,1.8){{\tiny \mbox{$k_F$}}}
\put(9.8,2.5){{\tiny \mbox{$0$}}}
\bezier{500}(8,5.5)(10,0)(12,5.5)
\put(9.8,1.3){(b)}
\end{picture}
\label{fig1}
\caption[Fig1]{%
(a) The Fermi surface consists of a single point
$k=k_F$.
(b) Schematic illustration for the
bosonization picture.
}%
\end{figure}

\end{document}